\documentclass[11pt]{article}
\usepackage{latexsym}  
\usepackage{amssymb}
\usepackage{graphicx}
\usepackage{amsmath}
\usepackage{color}

\begin{document}


\begin{center}
{\Large\bf  Tests of a Family Gauge Symmetry Model at $10^3$ TeV Scale}
  %
  
\vspace{5mm}

{\bf Yoshio Koide$^{a}$, Yukinari Sumino$^{b}$ and 
Masato Yamanaka$^{c}$}

{\it (a) Department of Physics, Osaka University,  
Toyonaka, Osaka 560-0043, Japan} \\

E-mail address: koide@het.phys.sci.osaka-u.ac.jp

{\it (b) Department of Physics, Tohoku University,  
Sendai 980-8578, Japan} \\

E-mail address: sumino@tuhep.phys.tohoku.ac.jp

{\it (c) MISC, 
Kyoto Sangyo University,  Kyoto 603-8555, Japan} \\

E-mail address: yamanaka@cc.kyoto-su.ac.jp

\end{center}

\begin{abstract}
Based on a specific model 
with U(3) family gauge symmetry at $10^3$ TeV scale,
we show its experimental signatures to search for.
Since the gauge symmetry is introduced with a special purpose, 
its gauge coupling constant and gauge boson mass spectrum 
are not free.  
The current structure in this model leads to family number 
violations via exchange of extra gauge bosons. 
We investigate present constraints from flavor changing 
processes and discuss
visible signatures at LHC and lepton colliders.
\end{abstract}



\vspace{3mm}


\section{Introduction}

In the current flavor physics, it is a big concern whether
flavors can be described by concept of ``symmetry" or not.
If the flavors are described by a symmetry (family symmetry), 
it is also
interesting to consider that the symmetry is gauged. 
(For an earlier work of gauge SU(3) symmetry, for example, 
see Ref.\cite{Yanagida:1979gs}.)
Most models with a family gauge symmetry have been introduced 
for the purpose of 
understanding mass spectra and mixings of quarks and leptons.
However, it is difficult to exclude such models by the present
and near future experiments, because in most models 
the gauge coupling constant $g_f$ and gauge boson masses
are free parameters.
In the present paper, 
we pay attention to a specific model with a U(3) family gauge 
symmetry which was proposed by one of the authors 
(YS)~\cite{Sumino09PLB,Sumino09JHEP}.
In contrast to the conventional U(3) family gauge model,
the present model has been introduced 
to explain the charged lepton spectrum with
high precision.
Therefore, the gauge coupling constant $g_f$ is fixed
with respect to the standard electroweak gauge coupling
constants as $g_f/2=e=g_2 \sin\theta_W$, and 
the mass spectrum of the gauge bosons is also fixed  
(see eq.(8) below).
As a result, we can give definite predictions, which
may allow these gauge bosons to be clearly 
detected or excluded in forthcoming 
experiments.

First, let us give a short review: 
``Why do we need a family gauge symmetry?"
In the charged lepton sector, 
we know that an empirical relation~\cite{Koidemass}
\begin{equation}
\begin{split}
   K \equiv 
   \frac{m_e +m_\mu + m_\tau}
   {(\sqrt{m_e} + \sqrt{m_\mu} + \sqrt{m_\tau})^2} 
   = \frac{2}{3} 
\end{split}   \label{K_relation}   
\end{equation} 
is satisfied with the order of $10^{-5}$ 
with the pole masses, i.e. $K^{pole}=(2/3)\times (0.999989 \pm
0.000014)$ \cite{PDG08}, while it is only valid with 
the order of $10^{-3}$ with the running masses, 
i.e. $K(\mu)=(2/3)\times (1.00189 \pm 0.00002)$ 
at $\mu =m_Z$. 
In conventional mass matrix models, ``mass" means not 
``pole mass" but ``running mass." 
Why is the mass formula (\ref{K_relation}) so remarkably satisfied with  
the pole masses? 
This has been a mysterious problem as to the relation 
(\ref{K_relation}) for long years.
Recently, a possible solution to this problem has been 
proposed by one of the authors 
(Y.S.)~\cite{Sumino09PLB,Sumino09JHEP}:
The deviation of $K(\mu)$ from $K^{pole}$ is 
caused by a logarithmic term $m_{ei}\log(\mu/m_{ei})$ 
in the running mass.
It was advocated that a family symmetry is gauged, and that
the logarithmic term in the 
radiative correction to $K(\mu)$
due to photon is  canceled by that due to  
family gauge bosons.
(This does not mean $m_{ei}(\mu)=m^{pole}_{ei}$.)
In order that
cancellation works correctly, the left-handed lepton 
field $\psi_L$ and 
its right-handed partner $\psi_R$ should be assigned to
${\bf 3}$ and ${\bf 3^*}$ of U(3) \cite{Appelquist06}, 
respectively, differently from 
the conventional assignment~\cite{Yanagida:1979gs} 
$(\psi_L, \psi_R) = ({\bf 3}, {\bf 3})$.


The assignment $(\psi_L, \psi_R)=({\bf 3}, {\bf 3^*})$ can 
induce interesting observable effects.   
In the conventional assignment, a family gauge 
boson $A_j^i$ couples to a current component  
$(J_\mu)_i^j = \bar{\psi}_L^j \gamma_\mu \psi_{Li} +
\bar{\psi}_R^j \gamma_\mu \psi_{Ri}$,
while in the present model, the gauge boson $A_j^i$ 
couples to 
\begin{equation}
\begin{split}
   (J_\mu)_i^j 
   = \bar{\psi}_L^j \gamma_\mu \psi_{Li}  
   - \bar{\psi}_{Ri} \gamma_\mu \psi_{R}^j .
\end{split}   \label{current_general}        
\end{equation}
In general, the currents (2) cause the violation of  
individual family number $N_f$ by $|\Delta N_f|=2$.
The influence of the family number violation is determined 
by the family gauge coupling constant $g_f$ and each 
family gauge boson mass $m_{fij} \equiv m(A_i^j)$. 
Here, for simplicity, the family current structure has been 
presented by a field $\psi$ as a representative of 
quarks $u$ and $d$ and leptons $e$ and $\nu$. 
For example, the charged lepton current component $(J_\rho)_1^2$ is
given by
\begin{equation}
\begin{split}
   (J_\rho)_1^2 
   = \bar{\mu}_L \gamma_\rho e_{L} 
   - \bar{e}_{R} \gamma_\rho \mu_{R} .
\end{split}   \label{current_mue}   
\end{equation}
This causes an $e$ (or $\mu$) lepton-number-violating process
$e^-+e^- \rightarrow \mu^- +\mu^-$ through the effective 
current-current interaction
\begin{equation}
\begin{split}
   {\cal L}^{eff} 
   = \frac{G_{f12}}{\sqrt2} \left[ 
   \bar{\mu} \gamma_\rho (1 -\gamma_5) e \right]
   \left[ \bar{\mu} \gamma^\rho (1+\gamma_5) e \right] +h.c., 
\end{split}   \label{Leff1}         
\end{equation}
where ${G_{f12}}/{\sqrt2} ={g_f^2}/{8 (m_{f12})^2}$
($m_{f12}=m(A_2^1)$).

In order to realize the cancellation mechanism 
between photon and family gauge bosons, 
$g_f$ should be related to the electric charge $e$ as
\begin{equation}
\begin{split}
   \frac{1}{4} g_f^2 = e^2 \equiv g_2^2 \sin^2 \theta_W ,
\end{split}   \label{gf}    
\end{equation}
where $g_2$ is the gauge coupling constant of SU(2)$_L$.
In \cite{Sumino09PLB,Sumino09JHEP} a speculation is given 
that the relation (\ref{gf}) 
may originate from unification of SU(2)$_L$ 
and family U(3) gauge symmetries at $10^2$--$10^3$~TeV scale;
the level of tuning of the unification scale required in this
scenario is estimated to be a factor of 3 to match
the present experimental accuracy of eq.~(\ref{K_relation}).
This model of charged lepton sector has been constructed
in the context of an effective field theory
with a cut--off scale $\Lambda\sim 10^3$--$10^4$~TeV, 
assuming this unification scenario and incorporating the
family U(3) gauge symmetry.
The masses of $A_i^j$ are predicted to be in the
$1 - 1000$~TeV range.

Thus, the ratio of the coefficients of the 
four-Fermi contact interactions 
is given by
\begin{equation}
\begin{split}
   \frac{G_{fij}}{G_F} 
   = 4 \sin^2 \theta_W 
   \left( \frac{m_W}{m_{fij}} \right)^2 
   = \frac{5.98\times 10^{-3}}
   {(m_{fij}\, {\rm [TeV}])^2} .
\end{split}   \label{G}           
\end{equation}
Here $G_{fij}/\sqrt{2} = g_f^2/8 m_{fij}^2$ and 
$G_F/\sqrt{2} = g_2^2/8 m_W^2$. 
In this model, Yukawa coupling constants $Y_e^{eff}$ of 
the charged leptons are effectively given by
\begin{equation}
\begin{split}
   (Y_e^{eff})_{ij} 
   = \frac{1}{\Lambda^2} \sum_{a=1}^3 
   \langle (\Phi_e)_{ia} \rangle
   \langle (\Phi_e^T)_{aj} \rangle ,
\end{split}   \label{Yeff}           
\end{equation}
where $\Phi_e$ is a scalar with $({\bf 3},{\bf 3})$ of
family U(3)$\times$O(3) symmetries.
(Here,  the family U(3)$\times$O(3) symmetries 
originate from a U(9) family symmetry \cite{Sumino09JHEP}, and
only U(3) gauge symmetry can contribute to the radiative
correction of the running masses of charged leptons 
below the cut--off scale $\Lambda$, 
at which the charged lepton 
mass relation (1) is given exactly.)
In other words, the VEV matrix $\langle \Phi_e \rangle$
is given as 
$\langle \Phi_e \rangle = {\rm diag}(v_1,v_2,v_3) 
\propto {\rm diag}(\sqrt{m_e}, \sqrt{m_\mu}, \sqrt{m_\tau})$.
[A prototype of such an idea for the charged lepton 
masses is found in Ref.~\cite{K-mass90} related to
the mass formula (\ref{K_relation}).]
Then, the gauge symmetry U(3) is completely broken by
$ \langle \Phi_e \rangle \neq 0$, so that the 
gauge boson masses $m_{fij}$ are related to
the charged lepton masses as~\cite{Sumino09JHEP}
\begin{equation}
\begin{split}
   (m_{fij})^2 \equiv m^2(A_i^j) \propto m_{ei} + m_{ej} .
\end{split}   \label{mf}                     
\end{equation}
The mass spectrum (8) is essential in this model.
For example, if we assume $(Y_e^{eff})_i^j \propto \sum_k 
\langle (\Phi_e)_i^k\rangle \langle (\Phi_e)_k^j\rangle$,
we cannot obtain the relation (8).
It is assumed that other scalar VEV's with non-zero family
charge, if they exist,
have much smaller magnitudes
than $ \langle \Phi_e \rangle$, such that they do not
affect the family gauge boson spectrum.
This is crucial to protect the cancellation mechanism
within the present scenario.

The purpose of the present paper is to discuss how to test
this family gauge symmetry
within the above model.
We note that this model is incomplete, e.g.\ the quark and
neutrino sectors are not included, anomaly of the
family gauge symmetry is not canceled.\footnote{
Above the scale of the family symmetry breaking
the gauge anomaly should cancel.
We assume existence of such a more complete model, in which
all the fermions except the Standard-Model fermions
acquire masses of the order of the symmetry
breaking scale ($\sim m_{fij}$) and decouple
from the low energy spectrum.
}
We focus only on the family gauge interactions,
which are fairly independent of the details of the model. 
We examine the interactions with 
$|\Delta N_f|=2$ via the gauge boson $A_2^1$.
In the next section, we estimate a lower bound of its mass 
$m_{f12}$ from the experimental limit on the branching ratio 
of a rare kaon decay $K^+ \rightarrow \pi^+ \mu^- e^+$,
assuming that the quarks are assigned
to multiplets of the $U(3)\times O(3)$ family gauge group in the same
way as the charged leptons.\footnote{
This is the only (minimalistic) assumption we impose on top of the
original model \cite{Sumino09JHEP}.
}
We also discuss $K^0$-$\bar{K}^0$ mixing and muonium into 
antimuonium conversion.
(For a review of searches for signatures with 
$|\Delta N_f|=2$, see, for example, Ref.\cite{Kuno-Okada01}.)
In Sec.~3, we investigate possible signatures in collider 
experiments, such as $e^- + e^- \rightarrow \mu^- + \mu^-$ 
production.
Since the mass of the lightest gauge boson $A_1^1$ 
may take a value within 1-- 10 TeV range,  
we may expect a production $p+p \rightarrow 
A_1^1 + X \rightarrow (e^+ e^-)+X$ at LHC.
We estimate the production cross section and decay rate.
Finally, Sec.~4 is devoted to a summary.


\section{Lower bounds for the gauge boson masses} 
 
First, in order to see more details of the characteristic 
current structure (\ref{current_general}), we discuss 
the flavor changing neutral currents relevant for $\mu$ and 
$e$. 
According to eq.~(\ref{current_general}), the current $(J_\rho)_1^2$ 
can be written as  
\begin{equation}
\begin{split}
   (J_\rho)_1^2 
   = \bar{\mu}_L \gamma_\rho e_L
   - \bar{e}_R \gamma_\rho \mu_R
   =  (J_V)_\rho - (J_A)_\rho ,
\end{split}   \label{aa}    
\end{equation}
where $(J_V)_\rho =({1}/{2}) (\bar{\mu}\gamma_\rho e 
-\bar{e} \gamma_\rho \mu)$ and $(J_A)_\rho =({1}/{2}) 
 (\bar{\mu}\gamma_\rho \gamma_5 e +
\bar{e} \gamma_\rho \gamma_5 \mu)$.
The vector current $J_V^\rho$ and axial current $J_A^\rho$
have $CP=-1$ and $CP=+1$, respectively. 
However, this does not mean that the effective current-current 
interactions cause $CP$-violating interactions. In fact, the current 
$(J_\rho)_2^1$ is written as $(J^\rho)_2^1 
= \bar{e}_L \gamma^\rho \mu_{L} - \bar{\mu}_{R} \gamma^\rho e_{R}      
   = - (J_V)^\rho - (J_A)^\rho$, 
so that the effective current-current interaction 
is $CP$ conserving:
\begin{equation}
\begin{split}
   {\cal L}^{eff} 
   = 4 \frac{G_{f12}}{\sqrt2} (J_\rho)^2_1(J^\rho)_2^1  
   = -4 \frac{G_{f12}}{\sqrt2} 
   \Big[ (J_V)_\rho (J_V)^\rho - (J_A)_\rho (J_A)^\rho \Bigr] .
\end{split}    \label{Leff2}  
\end{equation}

Next we discuss rare kaon decays. 
Note that, in this model, the family number $i=(1,2,3)$ 
is defined as $(e_1,e_2,e_3)=(e,\mu,\tau)$ in the charged 
lepton sector. 
If we assume $(d_1,d_2,d_3) \simeq (d,s,b)$ in the down-quark 
sector, the gauge boson masses $m_{f12}$ can be constrained
by the rare kaon decay searches.  
In general, a down-quark mass matrix $M_d$ is not necessarily
diagonal in the diagonal basis of the charged lepton mass 
matrix $M_e$.
For simplicity,  we assume that $M_d$ is Hermitian and 
consider only a $d$-$s$ mixing 
\begin{equation}
\left( \begin{array}{c}
d_0 \\
s_0 \\
b_0 
\end{array} \right) = U_d \left( \begin{array}{c}
d \\
s \\
b 
\end{array} \right) = \left( \begin{array}{ccc}
\cos\theta & -\sin\theta & 0 \\
\sin\theta & \cos\theta & 0 \\
0 & 0 & 1
\end{array} \right)  \left( \begin{array}{c}
d \\
s \\
b 
\end{array} \right),   
\label{mixingmat1}
\end{equation} 
where 
the down-quark mass matrix $M_d$ is given in the flavor basis 
in which the charged lepton mass matrix $M_e$ is diagonal, and
$M_d$ is diagonalized as $U_d^\dagger M_d U_d = 
{\rm diag}(m_d, m_s, m_b)$.
In this case, the down-quark current $(J_\mu^{(d)})_1^2$ is given by
\begin{equation}
\begin{split}
   &(J_\mu^{(d)})_1^2 
   = \bar{s}_L^0 \gamma_\mu d^0_L 
   - \bar{d}^0_R \gamma_\mu s^0_R     \\
   &~= \frac{1}{2}(\bar{s}\gamma_\mu d - \bar{d}\gamma_\mu s)
   - \frac{1}{2}(\bar{s}\gamma_\mu \gamma_5 d 
   + \bar{d}\gamma_\mu \gamma_5 s) \cos 2\theta     
   + \frac{1}{2}(\bar{s}\gamma_\mu \gamma_5 s 
   - \bar{d}\gamma_\mu \gamma_5 d) \sin 2\theta ,
\end{split}   \label{Jd}          
\end{equation}
where the first, second and third terms have $CP=-1$, 
$+1$ and $+1$, respectively.
Note that the vector current is independent of the 
mixing angle $\theta$.
(However, this is valid only 
with the mixing matrix eq.~(\ref{mixingmat1}).)

As an example of the $s$-$d$ current, let us discuss
a decay of neutral kaon into $e^{\pm}+\mu^{\mp}$.
In eq.~(\ref{Jd}), only the second term is relevant to
a neutral kaon with
spin-parity $0^-$, which
has $CP=+1$.
Since the observed neutral kaons $K_S$ and $K_L$ have
$CP=+1$ and $CP=-1$, respectively, in the limit of 
$CP$ conservation, we must identify the second term
in eq.~(\ref{Jd}) as $K_S$ (not $K_L$).
Hence, a stringent lower limit of
$m_{f12}$ cannot be extracted from the present 
experimental limit \cite{PDG08} 
$BR(K_L \rightarrow e^{\pm} \mu^{\mp})<4.7 \times 10^{-12}$.

Instead, the lower limit of $m_{f12}$ can be 
obtained from the rare kaon decays $K^+ \rightarrow \pi^+ 
+ e^{\pm} +\mu^{\mp}$.
The $K \rightarrow\pi$ decay is described by the first term (vector 
currents) in eq.~(\ref{Jd}), which can be replaced by 
$i(\pi^- \stackrel{\leftrightarrow}{\partial}_\rho K^+)$.
Hence, 
\begin{equation}
\begin{split}
   {\cal L}^{eff} 
   &= 2 (G_{f12}/\sqrt{2}) (\bar{s} \gamma_\rho d) 
   (\bar{e} \gamma^\rho \mu - \bar{\mu} \gamma^\rho e)    \\
   &\Rightarrow 2 (G_{f12}/\sqrt{2}) i(\pi^-
   \stackrel{\leftrightarrow}{\partial}_\rho K^+)
   (\bar{e} \gamma^\rho  \mu - \bar{\mu} \gamma^\rho e) .
\end{split}   \label{Leff3}    
\end{equation}
Since the effective interaction for $K^+ \rightarrow \pi^0 \mu^+
\nu_\mu$ is given by
${\cal L}_{weak} =  ({g_2^2}/{2m_W^2}) V_{us} 
(\bar{s}_L \gamma_\rho u_L) (\bar{\mu}_L \gamma^\rho \nu_{\mu L})$,
the ratio $BR(K^+ \rightarrow \pi^+ e^\pm \mu^\mp)/
BR(K^+ \rightarrow \pi^0 \mu^+ \nu_\mu)$ is given by
\begin{equation}
\begin{split}
   R 
   = \frac{\bigl[ 2 \cdot (G_{f12}/\sqrt{2}) \bigr]^2}
   {2 |V_{us}|^2 (1/\sqrt{2})^2 (G_F/\sqrt{2})^2} 
   = 67.27 \left(\frac{m_W}{m_{f12}}\right)^4 ,
\end{split}   \label{R}        
\end{equation}
in the approximation $m(\pi^+)=m(\pi^0)$ and 
$m(e^-)=m(\nu_\mu)=0$.
The present experimental limits \cite{PDG08} 
$BR(K^+ \rightarrow \pi^+ e^- \mu^+) <1.3 \times 10^{-11}$
and 
$BR(K^+ \rightarrow \pi^+  \mu^- e^+) <5.2 \times 10^{-10}$
together with $BR(K^+ \rightarrow \pi^0 \mu^+ \nu_\mu)=
(3.35\pm 0.04)\times 10^{-2}$ give lower limits of
the gauge boson mass $m_{f12}$ as shown in Table 1.
Note that the mode $K^+ \rightarrow \pi^+ e^+ \mu^-$
has $|\Delta N_f|=2$, which we are interested in,
while the mode $K^+ \rightarrow \pi^+ e^- \mu^+$ has 
$|\Delta N_f|=0$. 
We can estimate lower bounds of other gauge boson masses, 
$m_{f11}$, $m_{f13}$, etc., from the lower bounds 
of $m_{f12}$ using the relation (8).
The results are listed in Table 1. 
In the present model, the mass $m_{f33}$ of the 
heaviest gauge boson $A_3^3$ is predicted in the 
$10^2$--$10^3$~TeV range.
On the other hand, the lower bound
of $m_{f33}$ estimated from $K^+ \rightarrow \pi^+ e^- \mu^+$
is 300~TeV as seen in Table 1. 
Therefore, the lower bound of each gauge boson listed in
Table 1 seems to be almost near to its upper bound. 
In other words, the mass values given in Table 1 suggest 
that experimental observations 
of family gauge boson effects soon become within our reach. 
If we consider, however, a more general mixing of 
the down-type quarks,
we obtain suppression factors to the above branching
ratios.
In this case, constraints to the gauge boson masses
become looser.

A constraint on $m_{f12}$ can also be obtained from the 
observed value of the $K^0$-$\bar{K}^0$ mixing. 
The prediction for the $K^0$-$\bar{K}^0$ mixing
in the present model is more sensitive to
the mixing of the down-type quarks than for the rare
kaon decays.
Even with the simple ansatz eq.~(\ref{mixingmat1}),
the prediction depends on the value of $\theta$.
Hence, first we present the prediction in the no-mixing case
($\theta=0$) as a reference for small mixing,
and afterwards
we discuss the case with a general down-type quark mixing.
In contrast to the $(V-A)(V-A)$-type effective interaction
$[\bar{s} \gamma_\mu (1-\gamma_5) d]
[\bar{s} \gamma^\mu (1-\gamma_5) d]$
induced in conventional models,
the present model induces the $(V-A)(V+A)$-type
effective interaction
$[\bar{s} \gamma_\mu (1-\gamma_5) d]
[\bar{s} \gamma^\mu (1+\gamma_5) d]$.
This leads to the $K^0$-$\bar{K}^0$ mixing
\begin{equation}
\left[2+\frac{4}{3} \left( \frac{m_K}{m_s+m_d} \right)^2 \right] 
\langle \bar{K}^0 |\bar{s} \gamma_\mu (1-\gamma_5) d |0 \rangle
\langle 0| \bar{s} \gamma^\mu (1-\gamma_5)d | K^0\rangle  
\label{K0K0barmixing}
\end{equation}
under the vacuum saturation approximation, which should be 
compared with
\begin{equation}
\frac{8}{3} \, \langle \bar{K}^0 |
\bar{s} \gamma_\mu (1-\gamma_5) d |0 \rangle
\langle 0| \bar{s} \gamma^\mu (1-\gamma_5) d 
| K^0\rangle   
\end{equation}
in the conventional case.
With eq.~(\ref{K0K0barmixing}) we find a lower bound for
$m_{f12}$ of order $10^3$~TeV, which serves as a reference for small
down-type quark mixing.
We note that this bound is much more stringent than the values
listed in Table~1 (although it may still not completely rule out
the model if we take into account uncertainties in the
estimate of the unification scale in the model).

If we take into account a general mixing of the down-type quarks,
the prediction for the $K^0$-$\bar{K}^0$ mixing can be
either larger or smaller.
In particular, in the case that the 
mixing matrices $U_{dL}$ and $U_{dR}$
are complex, without specific tuning of the matrices,
generally a very stringent constraint is imposed from 
the $CP$ violation in the $K^0$-$\bar{K}^0$ mixing:
$m_{f12} \gtrsim 10^5$~TeV \cite{Maehara:1979kf},
which rules out the present model.
On the other hand, 
there exists a parameter region
(parametrized by a set of continuous parameters), 
where the contribution to
the $K^0$-$\bar{K}^0$ mixing vanishes.
Even if we restrict the mixing matrices to real
(orthogonal) matrices, such solutions exist with
rather simple forms.
For instance, in the case $U_{dR}={\bf 1}$ and
\begin{eqnarray}
U_{dL}\in 
{\small
\left\{
\left(\begin{array}{ccc}
0 & \pm 1 & 0\\
c_\theta & 0 & -s_\theta\\
s_\theta & 0 & c_\theta
\end{array}\right),~
\left(\begin{array}{ccc}
0 &  -s_\theta & c_\theta\\
\pm 1 & 0 & 0\\
0 & c_\theta &  s_\theta
\end{array}\right),~
\left(\begin{array}{ccc}
c_\theta & 0 & -s_\theta\\
s_\theta & 0 & c_\theta\\
0 & \pm 1 & 0\\
\end{array}\right),~
\left(\begin{array}{ccc}
0 & 0 & \pm 1\\
0 & \pm 1 & 0\\
\pm 1 & 0 & 0\\
\end{array}\right)
\right\},
}
\nonumber\\ &&
\label{solmixingmat}
\end{eqnarray}
($s_\theta\equiv\sin\theta,c_\theta\equiv\cos\theta$ for 
$\forall \theta$),
the induced four-Fermi operator for the $K^0$-$\bar{K}^0$ mixing
vanishes 
due to the characteristic form of the family gauge 
interactions.\footnote{
Another example of solutions is $U_{dL}={\bf 1}$ and $U_{dR}$ of
the form given in eq.~(\ref{solmixingmat}).
}
In general (but restricting to orthogonal mixing matrices
to circumvent constraints from the $CP$ violation), 
if the mixing induces a coupling of
the $d$-$s$ current to the lightest gauge boson $A^1_1$,
the $K^0$-$\bar{K}^0$ mixing tends to be more
enhanced and the bounds
for the gauge boson masses tend to be severer.
For certain choices of the mixing matrices, 
[e.g.\ $U_{dR}$ sufficiently close to ${\bf 1}$ and $U_{dL}$ to
eq.~(\ref{solmixingmat})],
the induced four-Fermi operators are suppressed,
and the lower bound for $m_{f12}$ can be reduced
much below $10^3$~TeV.

Let us briefly discuss bounds from the 
observed $D^0$--$\bar{D}^0$
mixing.
In order to predict 
contributions of family gauge
boson exchanges to the $D^0$--$\bar{D}^0$
mixing, we need to know the mixing matrices for the
up-type quarks $U_{uL}$ and $U_{uR}$.
Of these,
$U_{uL}$ is related to $U_{dL}$ by
$V_{\rm CKM}=U^\dagger_{uL} U_{dL}$, where 
$V_{\rm CKM}$ is the Cabibbo-Kobayashi-Maskawa (CKM) matrix,
while $U_{uR}$ is unknown.
Naively the lower bound from the $D^0$--$\bar{D}^0$
mixing on $m_{f12}$
is of order $10^2$--$10^3$~TeV.
Since the constraint on $CP$ violation is at present
not very tight, the bounds on the $CP$ phases
in $U_{uR}$ are not very demanding.
On the other hand, for $U_{uL}$ corresponding to
$U_{dL}$ of eq.~(\ref{solmixingmat}), there always exist
$U_{dR}$ which suppress 
the induced four-Fermi operator for the $D^0$-$\bar{D}^0$ mixing,
although we have not found particularly simple forms for
the combination $U_{uL}$ and $U_{uR}$.\footnote{
This is partly due to the fact that we do not know what can
be regarded as ``simple'' forms, given the constraint 
$V_{\rm CKM}=U^\dagger_{uL} U_{dL}$ by the 
present experimental data.
}
We present a detailed analysis of the effects of the
quark mixing in our future work.

We also note that if the CKM quark mixing originate from
VEV's of scalar fields (with non-trivial
$U(3)$ charges) 
other than $\langle \Phi_e \rangle$, in general they may contribute to
mixings of family gauge bosons, and therefore they would receive a tight
constraint from the experimental data for the $K^0$--$\bar{K}^0$
mixing.
This is, however, highly dependent on the model of the quark 
sector, in comparison to the constraints analyzed above.\footnote{
Introduction of other $U(3)$-breaking 
scalar VEV's is {\it not} mandatory for generating CKM quark mixing.
For instance, quark mass matrix can be generated from
$\Phi_e S_q \Phi_e^T$, where $S_q$ has only $O(3)$ charge and off-diagonal; 
this form is similar to the lepton mass matrix of the present model.
[$S_q$ may even have a non-trivial $CP$-phase, since 
$U(3)\times O(3)$ is embedded into $U(9)$.]
}

We summarize here our standpoint with respect to 
the constraints on the gauge boson mass
from the quark sector, namely from 
the charged kaon decays, $K^0$-$\bar{K}^0$ mixing, 
and $D^0$-$\bar{D}^0$ mixing.
The severe constraint from the $CP$ violation in 
the $K^0$-$\bar{K}^0$ mixing shows that
$CP$ phases in the down-type quark mixing $U_{dL}$ and
$U_{dR}$ are absent or do not contribute to the $K^0$-$\bar{K}^0$ mixing,
for the model to be viable.
A simple possibility is to constrain $U_{dL}$ and
$U_{dR}$ to be real, and this will be assumed in the
rest of our analysis.
The
constraints from the $K^0$-$\bar{K}^0$ mixing
and $D^0$-$\bar{D}^0$ mixing indicate that
$m_{f12}\gtrsim 10^3$~TeV, without tuning of the mixing matrices.
These bounds, however, can be lowered to order $10^2$~TeV
(roughly the expected size of this family gauge
boson mass)
in a non-negligible region of the parameter space
of the mixing matrices.
In order to reduce $m_{f12}$ to a much lower mass range,
naively it seems to require considerable fine tuning of the mixing
matrices.
Nevertheless, given the simple forms of the down-type quark
mixing eq.~(\ref{solmixingmat}), we may as well  keep our
mind open for a possibility that Nature indeed conspires to
realize such a case.

\begin{table}  
\caption{{Masses of the gauge bosons $A_1^1$, 
$A_2^1$, $A_3^1$ and $A_3^3$, and their lower bounds 
from rare kaon decays,
assuming the down-type quark mixing
eq.~(\ref{mixingmat1}).
Their relative sizes are also shown.  }} 
\begin{center}
\begin{tabular}{lllll} \hline
  
  & $m_{f11}$  
  & $m_{f12}$ 
  & $m_{f13}$ 
  & $m_{f33}$ 
  \\ \hline
  Relative sizes
  & $\sqrt{2m_e}$ 
  & $\sqrt{m_\mu +m_e}$
  & $\sqrt{m_\tau +m_e}$ 
  & $\sqrt{2m_\tau}$
  \\
  
  & $0.0981127$ 
  & $1.00000$ 
  & $4.09154$ 
  & $5.78448$
  \\ \hline
  $K^+ \rightarrow \pi^+ \mu^- e^+$
  & 2.1 TeV 
  & 21 TeV 
  & 86 TeV 
  & 120 TeV
  \\
  $K^+ \rightarrow \pi^+ e^- \mu^+$
  & 5.1 TeV 
  & 52 TeV 
  & 210 TeV 
  & 300 TeV
  \\
  \hline
\end{tabular}
\end{center}
\end{table}  


%
As seen above, the bounds for $m_{f12}$
extracted from the quark sector are quite dependent on
the structure of the quark mixing matrices.
By contrast, a strict bound can be extracted from
a purely leptonic process independently of the quark sector,
since the interactions of the charged leptons with the
family gauge bosons are completely fixed.
In passing, let us comment on the leptonic processes $\mu \to 3\,e$ and
$\mu\to e\gamma$.
The effective interaction (\ref{Leff2}) include only  
$|\Delta N_f|=0$ and $|\Delta N_f|=2$ terms, 
whereas these processes have $|\Delta N_f|=1$.
Hence, these processes can occur only through family mixing in 
quark loops.
They are dependent on the quark mixing matrices; 
furthermore, the
constraints from these processes are looser than other
quark-mixing dependent ones
which we considered above.
Therefore, we do not discuss $\mu \to 3\,e$ and
$\mu\to e\gamma$ any further.
Here, we consider the muonium into antimuonium conversion $M(\mu^+ e^-) 
\rightarrow \overline{M}(\mu^- e^+)$, which has $|\Delta N_f|=2$.
The total $M\overline{M}$ conversion probability $P_{M\overline{M}}(B)$
under an external magnetic field $B$ 
is given by
$P_{M\overline{M}}(B) = {\delta^2}/{2
[\delta^2+ (E_M-E_{\overline{M}})^2 +\lambda^2]}$,
where $E_M$ and $E_{\overline{M}}$ are the energies of 
$M$ and $\overline{M}$, respectively, 
$\lambda$ is the bound muon decay width, and $\delta$ is 
defined by $\langle M |H_{M \overline{M}}|\bar M \rangle$ which is
proportional to $(G_{f12}/\sqrt2)/\pi a^3$ ($a$ is the electron 
Bohr radius).  
Here, the effective interaction describing $M\overline{M}$ 
conversion is given by eq.(4).  
This has the same $(V-A)(V+A)$ form as
the one corresponding to a dilepton model~\cite{dilepton},
and the formulation in this case
has been investigated by Horikawa and 
Sasaki~\cite{Horikawa96} in detail. 
It predicts 
$P_{M\overline{M}}(0) \simeq ({3}/{2}) {\delta^2}/{\lambda^2}$ and 
$\delta = -8 ({G_{f12}}/{\sqrt2}) ({1}/{\pi a^3})$.
It follows that
\begin{equation}
\begin{split}
   P_{M\overline{M}}(0) 
   =1.96 \times 10^{-5} \times \left( \frac{G_{f12}}{G_F} \right)^2   
   = \frac{7.01\times 10^{-10}}{\bigl( m_{f12} ~ [\text{TeV}] \bigr)^4} .
\end{split}   \label{P}    
\end{equation}
For example, for $m_{f12}=21$ TeV and $52$ TeV, eq.~(\ref{P}) 
predicts $P_{M\overline{M}}(0)=3.6\times 10^{-15}$ and
$9.6\times 10^{-17}$, respectively. 
Present experimental limit~\cite{Willmann99} of the total conversion
 probability integrated over all decay times is 
$P_{M\overline{M}}(B) \le 8.3 \times 10^{-11}$ (90\% CL) for $B=0.1$~T.
Since $S_B(0.1 {\rm T})=0.78$ for the case of $(V-A)(V+A)$ 
\cite{Horikawa96}, where $S_B(B)$ is defined by 
$P_{M\overline{M}}(B)=P_{M\overline{M}}(0) S_B(B)$, 
this bound leads to 
$P_{M\overline{M}}(0) \le 1.06 \times 10^{-10}$, and to
$G_{f12}/G_F \le 2.3 \times 10^{-3}$.
Thus, the lower bound of $m_{f12}$ is given by
\begin{equation}
\begin{split}
   m_{f12} \ge 20\, m_W = 1.6 \ {\rm TeV} .
\end{split}   \label{mf_bound}   
\end{equation}
This constraint is looser than the constraints
listed in Table~1 or from the $K^0$-$\bar{K}^0$/$D^0$-$\bar{D}^0$ mixing.
However, since the down-quark mixing matrices 
$U_{dL}$ and $U_{dR}$ are unknown at present apart 
from the CKM matrix,  
we would like to emphasize the importance of
observations in the pure
leptonic processes, independently of 
the bounds from the rare kaon decays.
In this respect, we expect that future experiments will
improve the bounds given in eq.~(\ref{mf_bound}).


\section{Search for signatures at collider experiments}

Next, we investigate possible signatures of the current-current
interaction with $|\Delta N_f|=2$ at collider experiments. 
Although a top-top production at LHC (via $u+u \rightarrow t+t$)
is very attractive, the cross section $\sim 10^{-6}$ pb at 
$\sqrt{s}=14$ TeV and for $m_{f13}=10^2$ TeV
would be too small to detect the signal.  
The cross section for
$e^- + p \rightarrow \mu^- + X$ amounts to
$\sigma \sim 10^{-5}$ pb at $E_p = 7$ TeV and
$E_e=400$ GeV for $m_{f12}=50$ TeV, which would 
also be difficult to detect, because of a large background 
$e^- + p \rightarrow \mu^- + \nu_e + \bar{\nu}_\mu + p$ 
with $\sigma \sim 10^{-1}$~pb.

The most clean reaction with $|\Delta N_f|=2$ is $e^- + e^- 
\to \mu^- + \mu^-$.  This reaction is expected at an optional 
experiment at a future $e^+ e^-$ linear collider. 
The current structure in this model shows that this reaction 
takes place  only between invertedly polarized 
electron pairs $e_\text{L}^- e_\text{R}^-$. 
This aspect is useful
for discriminating this model from others using
the polarized $e^-$ beams.
We obtain the differential cross section
\begin{equation}
\begin{split}
   \frac{d \sigma}{d\cos\theta} 
   = \frac{2 \pi \alpha_\text{EM}^2}{m_{f12}^4} s (1+\cos^2\theta) ,
\end{split}   \label{cross} 
\end{equation} 
and the total cross section $\sigma(e_\text{L}^- e_\text{R}^- 
\rightarrow \mu^- \mu^-) = (16 \pi \alpha_\text{EM}^2/3 m_{f12}^4) s$. 
Fig.~1 shows the {differential} cross sections 
$d\sigma(e_\text{L}^- 
e_\text{R}^- \to \mu^- \mu^-)/d\cos\theta$ at the c.m.\ energy 
$\sqrt{s} = 2$~TeV. 
The value of the family gauge boson mass $m_{f12}$ corresponding
to each line is displayed in the figure.
For $m_{f12} = 21$ TeV (52 TeV) and at
$\sqrt{s} = 2$~TeV, the total cross section is given by
$\sigma = 3.3 \times 10^{-2}$ ($8.7 \times 10^{-4}$)~fb. 
A high luminosity operation of a future lepton collider 
may lead to the model 
confirmation by observing the clean reaction with $|\Delta N_f| = 2$.

\begin{figure}[t!]
  \includegraphics[width=78mm,clip]{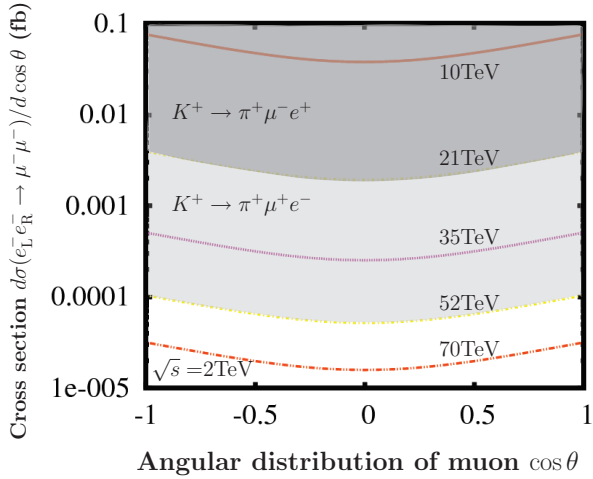}
  \caption{Differential cross section $d\sigma(e_\text{L}^- 
  e_\text{R}^- \to \mu^- \mu^-)/d\cos\theta$ vs.\ 
   $\cos\theta$.   We set $\sqrt{s} = 
  2$~TeV and $m_{f12} =$10, 35, and 70~TeV. 
  The light-shaded and
  dark-shaded regions represent the constraints
  from rare kaon decays listed in Table~1, which
  assume the down-type quark mixing eq.~(\ref{mixingmat1}).
  }
  \label{A12_ILC}
\end{figure}


\begin{figure}[t!]
  \includegraphics[width=80mm,clip]{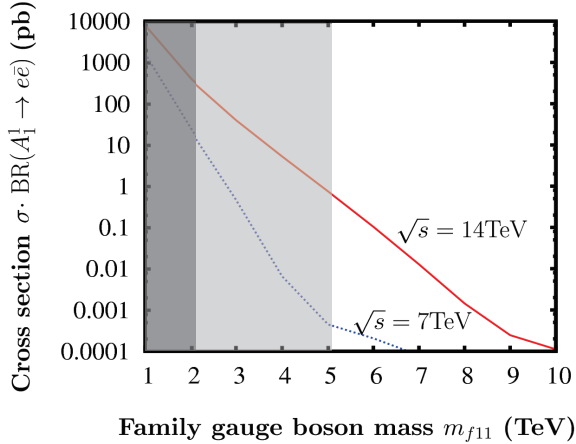}
  \caption{$\sigma (pp \to A_1^1 X) \cdot \text{BR}(A_1^1 \to 
  e^+ e^-)$ as a function of the family gauge boson mass $m_{f11}$. 
  The light-shaded (dark-shaded) region is the same as in Fig.~1.}
  \label{A11_LHC}
\end{figure}

\begin{table}  
\caption{Cross sections for the signal and Drell-Yan background, and
$S/\sqrt{N}$ corresponding to
an integrated
luminosity of 1~fb$^{-1}$, at LHC
$\sqrt{s} = 14$~TeV. No cuts are imposed.}
\begin{center}
\begin{tabular}{cclll} \hline
  &$m_{f11}$(TeV)~
  &signal (fb) ~~~~
  &DY BG (fb)   ~~
  &$S/\sqrt{N}$
  \\ \hline
  &2
  &$4.4 \times 10^{2}$
  &$1.6 \times 10^{-1}$
  &$1.1 \times 10^{3}$
  \\
  &3
  &$4.2 \times 10$
  &$1.5 \times 10^{-2}$
  &$3.4 \times 10^{2}$
  \\
  \hline
\end{tabular}
\end{center}
\end{table}

Finally, we discuss a search for the gauge boson $A_1^1$, 
which is the lightest one of the U(3) family gauge bosons.
For simplicity, we neglect the up-quark mixing as well as 
down-quark mixing, i.e. 
$(u_1,u_2,u_3)\simeq (u,c,t)$ and $(d_1,d_2,d_3)
\simeq (d,s,b)$.
The method is practically the same as that for 
$Z'$ boson. 
[For reviews of $Z'$, see, for instance, 
Refs.~\cite{Z-prime}.
In particular,
the highest limit of $Z'$ mass from direct searches
is about 1~TeV, which is much smaller than the
bounds on $m_{f11}$ in Table~1.] 
In conventional $Z'$ models, $Z'$ couples to fermions of
all flavors, whereas the $A_1^1$ boson couples only to the 
first generation, i.e., $A_1^1 \rightarrow e^+ e^-,
\nu_e \bar{\nu}_e, u \bar{u}, d \bar{d}$.
The total decay width and
the branching ratio 
are given, respectively, by 
\begin{equation}
\begin{split}
   &\Gamma(A_1^1\rightarrow all) 
   = (5/16\pi) g_f^2 m_{f11}
   = 5 \, \alpha_{em} m_{f11},
\\ &
   \text{BR} (A_1^1\rightarrow e^+ e^-) =  2/15,
\end{split}    \label{BRA11ee} 
\end{equation}
which are different from those of conventional $Z'$ models.
Since we presume that $A_1^1$ has a mass larger than
${\cal O}(1$~TeV),
it is not expected to find $A_1^1$ at Tevatron.
On the other hand, we may expect productions of $A_1^1$ at LHC.
In Fig.2, we show the cross section
$\sigma(p\, p\rightarrow A_1^1 \, X \rightarrow e^+ e^- X)
= \sigma(pp \rightarrow A_1^1 X) \cdot 
\text{BR} (A_1^1\rightarrow e^+ e^-)$
for $\sqrt{s}= 7$ TeV and 14 TeV. 
The cross sections are calculated with CalcHEP~\cite{Pukhov:2004ca} 
implementing eq.~(\ref{current_general}) and with the 
CTEQ6L code~\cite{Pumplin:2002vw} for the parton distribution function. 
When we reconstruct dilepton invariant masses $m(l^+ l^-)$, 
if we observe a peak in $m(e^+ e^-)$ but no peak in
$m(\mu^+ \mu^-)$, this will be a signal of
the new gauge boson $A_1^1$.
(This feature is unchanged even with up-quark mixing.)

The dominant backgrounds in the $A_1^1$ search,
after moderate event selection cuts,
are Drell-Yan
dielectrons
\cite{Aad:2009wy}.
Table~2 lists $S/\sqrt{N}$ as a measure of 
$A_1^1$ discovery reach for $m_{f11} \leq 3$~TeV.
Estimates of backgrounds within a window of
$\pm 4\Gamma_{Z'}\approx \pm \Gamma_{A^1_1}$ before any cut
are taken from \cite{Aad:2009wy}.
Comparing to the analysis given there, we anticipate
that, with an integrated luminosity of 10~fb$^{-1}$,
$m_{f11}$ up to several TeV would be within
discovery reach.
However, we leave a detailed study to our future work.


\section{Summary} 

At present, the cancellation mechanism based on U(3) 
family gauge symmetry is the only known one as 
a possible explanation for $K(\mu)=K^{pole}$.
Therefore, tests of the model are urgently required.

In this model, the family number $i=(1,2,3)$ is defined 
as $(e_1,e_2,e_3)=(e,\mu,\tau)$ in the charged lepton 
sector. 
Once we fix the mass matrix (or the mixing matrix) of
the down-type quarks in this basis,
we can extract constraints on the family gauge boson masses
from the rare kaon decay searches and from the observed
value of the $K^0$--$\bar{K}^0$ mixing.
Similarly if we fix the up-type quark mixing, we can extract
constraints from the $D^0$--$\bar{D}^0$ mixing.
The very stringent bounds from the $CP$ violation in the
$K^0$--$\bar{K}^0$ mixing rule
out contributions from $CP$ phases in the down-type quark
mixing matrices to this process.
Hence, we restrict our analysis to the real 
(orthogonal) down-type
quark mixing matrices.
Generally (without tuning of the mixing matrices)
we find $m_{f12}\gtrsim 10^3$~TeV from the
$K^0$--$\bar{K}^0$ and $D^0$--$\bar{D}^0$ mixing.
However, $m_{f12}\sim {\cal O}(10^2$~TeV) is also viable in a
non-negligible range in the parameter space of the mixing matrices,
which is consistent with the bounds from the rare
kaon decay searches.
We also find that, with certain simple forms of the
down-type quark mixing matrices, the contribution of
the family gauge bosons to the $K^0$--$\bar{K}^0$ mixing
vanishes.
Strictly speaking, if we allow for an arbitrary 
quark mixing, we cannot constrain the gauge boson masses
from these experimental data, since there exist
solutions, for which all these processes are
suppressed.
A quark-mixing independent bound is obtained from 
a purely leptonic process, muonium-antimuonium 
conversion, whose current lower bound reads $m_{f12}>1.6$~TeV.
More sensitive tests will come from
an upgrade of this experiment or from the process
$e_\text{L}^- e_\text{R}^- \to \mu^- \mu^-$ 
at ILC.  
Furthermore, if the lightest gauge boson $A^1_1$
happens to exist below several TeV,
we expect to observe a peak in $m(e^+ e^-)$ 
but no peak in $m(\mu^+ \mu^-)$ at LHC. 
These searches may uncover
an interesting possibility.

One may suspect that the bounds from
the $K^0$--$\bar{K}^0$ mixing
are too severe for the new physics signals to be observed at
LHC and/or ILC.
We note, however, that at present our knowledge on
the structure of the quark mixing matrices is rather limited,
and a conservative attitude would be to rely on
the current bounds from the purely leptonic process
$M(\mu^+e^-)$--$\overline{M}(e^+\mu^-)$.
In this regard, we stress that, although the
production rate of $A^1_1$'s at LHC depends on 
our assumption of the up- and down-quark mixing,
once they are produced, the family dependent
appearance of a peak
among the purely leptonic decay channels is independent
of the assumption and is a unique prediction of the
present model.
\\



\centerline{\large\bf Acknowledgments} 

The authors would like to thank Y.~Kuno for 
fruitful discussion.
Y.S.\ is grateful to H.~Yokoya for useful discussion.
Y.K. is also thank C.S.~Lim for a stimulating conversation 
about $K^0$-$\bar{K}^0$ mixing.  
The works of Y.K.\ and Y.S. are supported by JSPS 
(Nos.\ 21540266 and 20540246, respectively).
The work of M.Y.\ is supported by Maskawa Institute in Kyoto Sangyo 
University.


\end{document}